\documentclass{aa}
\usepackage{graphicx}
\begin{document}
   \title{
Effects of the core radius of an isothermal 
ellipsoidal gravitational lens on the caustics and 
the critical curves
} 

   \author{Hideki Asada
           \inst{1,2} 
          }

   \offprints{H. Asada\\ email: asada@phys.hirosaki-u.ac.jp}

   \institute{GReCO, Institute for Astrophysics at Paris, 
              98bis boulevard Arago, 75014 Paris, France 
        \and 
              Faculty of Science and Technology, 
              Hirosaki University, Hirosaki 036-8561, Japan} 

   \date{Received ; accepted }

   \abstract{
We study the effect of the core radius of an isothermal 
ellipsoidal gravitational lens on the caustics and 
the critical curves. 
We derive an analytic expression of the caustics for an isothermal 
ellipsoidal gravitational lens via a sixth-order algebraic equation. 
Since the expression is too long, by using another method 
we obtain a parametric representation of the critical curves 
in order to show analytically that there exist three cases: 
There are two curves for a small core radius, one 
for a quite large one, and no curves appear for an extremely large one, 
though the latter two cases are not realistic. The caustics are 
represented also by the same parameter.

   }

\authorrunning{Asada}

\titlerunning{Isothermal Ellipsoidal Gravitational Lens}

   \maketitle
%

\section{Introduction}
Gravitational lensing due to a galaxy is indispensable 
to probe its mass distribution and determine 
cosmological parameters (for reviews, Schneider et al. 1992). 
Cored isothermal ellipsoids are often used for 
modeling a galactic lens. 
Although such a model is quite simple, it enables us to understand 
many physical properties of the galactic lens. 
In addition, it fits with mass profiles implied 
by observations (e.g., Binney \& Tremaine 1987). 

The caustics are curves on the source plane perpendicular 
to the line of sight. 
They are important to understand the gravitational lensing, 
particularly when estimating its probability.  
Furthermore, the number of images changes 
if a source crosses the caustics. 
Similarly, critical curves located on the lens plane 
play a crucial role; giant arcs are observed near these 
critical curves. 

Analytic expressions for the caustics and critical curves 
were found for a {\it singular} isothermal ellipsoidal lens 
(Asada et al. 2003) during an algebraic study of 
the gravitational lens (Asada 2002, Asada et al. 2002). 
However, a {\it cored} isothermal ellipsoidal lens, 
which is more important and realistic than a singular one,  
seemed beyond an analytic approach, 
because the equation for a cored isothermal ellipsoidal lens 
is of much higher order and thus more complicated from 
an algebraic point of view (Asada et al. 2003). 
The existence of a core makes a crucial difference even if 
the core is small. 
For instance, the number of images for a singular isothermal 
ellipsoid is two or four, while it is one, three or five 
for a cored isothermal one (Asada et al. 2003). 
The main purpose of the present paper is 
to study the effect of the core radius on the caustics and 
the critical curves. 

This paper is organized as follows. 
In section 2, the lens equation for a cored isothermal ellipsoid 
is given in the form of a single sixth-order algebraic equation. 
The discriminant for this algebraic equation is also discussed.  
A parametric representation of the critical curves and caustics 
are obtained in section 3. 
Section 4 provides our conclusion. 

\section{Lens equation for an isothermal ellipsoidal lens} 
We consider an isothermal ellipsoidal lens with ellipticity 
$0\leq\epsilon< 1/5$ and the angular core radius $c$. 
The surface mass density projected onto the lens plane must be 
non-negative everywhere. This puts a constraint as $\epsilon<1$. 
A tighter constraint $\epsilon<1/5$ comes from the requirement 
that the density contour must be convex, which is reasonable 
for an isolated relaxed system. 
The lens equation is expressed as 
\begin{eqnarray}
\beta_1&=&\theta_1-\frac{(1-\epsilon)\theta_1}
{\sqrt{(1-\epsilon)\theta_1^2+(1+\epsilon)\theta_2^2+c^2}} , \\
\beta_2&=&\theta_2-\frac{(1+\epsilon)\theta_2}
{\sqrt{(1-\epsilon)\theta_1^2+(1+\epsilon)\theta_2^2+c^2}} ,  
\end{eqnarray}
where $\mbox{\boldmath$\beta$}=(\beta_1, \beta_2)$ and 
$\mbox{\boldmath$\theta$}=(\theta_1, \theta_2)$ denote 
the positions of the source and images, respectively. 

For simplicity, we introduce variables as 
$x\equiv\sqrt{1-\epsilon}\theta_1$, 
$y\equiv\sqrt{1+\epsilon}\theta_2$, 
$a\equiv\sqrt{1-\epsilon}\beta_1$ and 
$b\equiv\sqrt{1+\epsilon}\beta_2$, so that the lens equation 
can be rewritten as 
\begin{eqnarray}
a&=&x\left(1-\frac{1-\epsilon}{\sqrt{x^2+y^2+c^2}}\right) , 
\label{lenseq-cie1}
\\
b&=&y\left(1-\frac{1+\epsilon}{\sqrt{x^2+y^2+c^2}}\right) . 
\label{lenseq-cie2}
\end{eqnarray}

For off-axis sources, this lens equation is reduced to 
a single sixth-order algebraic one as (Asada et al. 2003)
\begin{eqnarray}
&&(a-x)^2
\Bigl[(x^2+c^2)[(1+\epsilon)a-2\epsilon x]^2
+(1-\epsilon)^2b^2x^2\Bigr] 
\nonumber\\
&&-(1-\epsilon)^2x^2[(1+\epsilon)a-2\epsilon x]^2
\nonumber\\
&&= 0 . 
\label{sixth-eqx}
\end{eqnarray} 
The other component is uniquely determined by 
\begin{equation}
y=\frac{(1-\epsilon)bx}{(1+\epsilon)a-2\epsilon x} . 
\label{y}
\end{equation}
It should be noted that the positions of the images 
can be found analytically for a source on a symmetry axis 
(Asada et al. 2003). 
However, it is sufficient to consider a case of off-axis sources 
to obtain expressions of the caustics and critical curves because 
these expressions are continuous functions of the source position.

By the discriminant for polynomial equations, we can determine 
a location where the number of real solutions changes 
(e.g., van der Waerden 1966), which corresponds to a change 
in the number of images in the context of the lensing. 
The discriminant for the sixth-order equation 
($\ref{sixth-eqx}$) is computed for instance by using 
the software Mathematica (Wolfram 2000). 
The result is a lengthy polynomial of 656 terms, 
which is not tractable. 
Hence we employ another method to obtain a parametric 
representation of the caustics and critical curves 
in a much shorter and thus more practical form.

\section{Critical curves and caustics} 
The critical curves are expressed as the vanishing of the Jacobian 
of the lens mapping, namely $\partial(a, b)/\partial(x, y)=0$. 
It becomes 
\begin{eqnarray}
&&(x^2+y^2+c^2)^2
-\Bigl(x^2+y^2+2c^2+e(x^2-y^2)\Bigr)
\nonumber\\
&&\times(x^2+y^2+c^2)^{1/2}+c^2(1-e^2)=0 . 
\label{critical}
\end{eqnarray}
This is rewritten as 
\begin{eqnarray}
&&(R+c^2)^2-(R+2c^2)(R+c^2)^{1/2}+c^2(1-e^2) 
\nonumber\\
&&=eR(R+c^2)^{1/2}\cos2\phi , 
\label{critical2}
\end{eqnarray}
where we introduced the polar coordinates 
$(x, y)=(r\cos\phi, r\sin\phi)$ and defined $R=r^2$. 

We will find allowed regions among $r$, $c$ and $e$, which is 
due to $-1\leq \cos2\phi \leq 1$ in Eq. ($\ref{critical2}$). 
In short, Eq. ($\ref{critical2}$) implies 
\begin{eqnarray}
&&(R+c^2)^2-(R+2c^2)(R+c^2)^{1/2}+c^2(1-e^2) 
\nonumber\\
&&\leq eR(R+c^2)^{1/2} , 
\label{critical2p}
\end{eqnarray}
and 
\begin{eqnarray}
&&(R+c^2)^2-(R+2c^2)(R+c^2)^{1/2}+c^2(1-e^2) 
\nonumber\\
&&\geq -eR(R+c^2)^{1/2} .  
\label{critical2m}
\end{eqnarray}

First, we consider Eq. ($\ref{critical2p}$), which is 
rewritten as 
\begin{eqnarray}
&&(R+c^2)^2+c^2(1-e^2) 
\nonumber\\
&&\geq 
\Big((R(1-e)+2c^2\Bigr)(R+c^2)^{1/2} . 
\end{eqnarray}
Since the right-hand side of this inequality is positive 
for $e<1$, this is recast in a polynomial form so that we can 
find the allowed region for $R$ in an algebraic manner. 
This polynomial expression is 
\begin{eqnarray}
F(R)&\equiv&\Bigl((R+c^2)^2+c^2(1-e^2)\Bigr)^2
\nonumber\\
&&-\Bigl(R(1-e)+2c^2\Bigr)^2(R+c^2) 
\nonumber\\
&\geq& 0 . 
\label{constraint1}
\end{eqnarray}
The discriminant $D_4$ of the fourth-order polynomial $F(R)$ 
is obtained (e.g. van der Waerden 1966) as 
\begin{eqnarray}
D_4&=&-27c^8(1+e)^4
\nonumber\\
&&\times
\Bigl((1-3e+3e^2-e^3)^2-c^4(1+e)^2\Bigr)^2 . 
\end{eqnarray}
This is non-positive, so that $F(R)=0$ has two real roots 
as $R=(1-e)^2-c^2$ and $c^{4/3}(1+e)^{2/3}-c^2$. 
We must find the smaller root satisfying 
a non-negativity of $R$.
In short, we compare between $0$, $(1-e)^2-c^2$ 
and $c^{4/3}(1+e)^{2/3}-c^2$. 
There exist four cases for Eq. ($\ref{constraint1}$), 
depending on $c$ as 
\begin{eqnarray}
\mbox{for} && c<\frac{(1-e)^{3/2}}{(1+e)^{1/2}} , 
\nonumber\\
&&0\leq R \leq c^{4/3}(1+e)^{2/3}-c^2 , 
\label{region-p-top}
\\
&&(1-e)^2-c^2\leq R , 
\\
\mbox{for} && \frac{(1-e)^{3/2}}{(1+e)^{1/2}}<c<1-e , 
\nonumber\\
&&0\leq R\leq (1-e)^2-c^2 , 
\\
&& c^{4/3}(1+e)^{2/3}-c^2\leq R , 
\\
\mbox{for} && 1-e<c<1+e , 
\nonumber\\
&& c^{4/3}(1+e)^{2/3}-c^2\leq R , 
\\
\mbox{for} && 1+e<c , 
\nonumber\\
&&0\leq R . 
\label{region-p-end}
\end{eqnarray}

Next, we consider Eq. ($\ref{critical2m}$), which implies 
\begin{eqnarray}
G(R)&\equiv&\Bigl((R+c^2)^2+c^2(1-e^2)\Bigr)^2
\nonumber\\
&&-\Bigl(R(1+e)+2c^2\Bigr)^2(R+c^2) 
\nonumber\\
&\leq& 0 , 
\label{constraint2}
\end{eqnarray}
whose discriminant $D_4'$ 
is obtained (e.g. van der Waerden 1966) as 
\begin{eqnarray}
D_4'&=&-27c^8(1-e)^4
\nonumber\\
&&\times
\Bigl((1+3e+3e^2+e^3)^2-c^4(1-e)^2\Bigr)^2 . 
\end{eqnarray}
This is non-positive, so that $G(R)=0$ has two real roots 
as $R=(1+e)^2-c^2$ and $c^{4/3}|1-e|^{2/3}-c^2$. 
We compare between $0$, $(1+e)^2-c^2$ 
and $c^{4/3}|1-e|^{2/3}-c^2$. 
We find three cases for Eq. ($\ref{constraint2}$) as 
\begin{eqnarray}
\mbox{for} && c<1-e , 
\nonumber\\
&&c^{4/3}|1-e|^{2/3}-c^2\leq R \leq (1+e)^2-c^2 , 
\label{region-m-top}
\\
\mbox{for} && 1-e<c<1+e , 
\nonumber\\
&&0\leq R\leq (1+e)^2-c^2 , 
\\
\mbox{for} && 1+e<c , 
\nonumber\\
&&\mbox{no allowed region for} \, R. 
\label{region-m-end}
\end{eqnarray}

Bringing together Eqs. ($\ref{region-p-top}$)-($\ref{region-p-end}$) 
and Eqs. ($\ref{region-m-top}$)-($\ref{region-m-end}$), 
we find four cases for Eqs. ($\ref{critical2p}$) 
and ($\ref{critical2m}$): 
\begin{eqnarray}
\mbox{For} && c<\frac{(1-e)^{3/2}}{(1+e)^{1/2}} , 
\nonumber\\
&&c^{4/3}|1-e|^{2/3}-c^2 \leq R \leq c^{4/3}(1+e)^{2/3}-c^2 , 
\label{region1}
\\
&&(1-e)^2-c^2 \leq R \leq (1+e)^2-c^2 , 
\\
\mbox{for} && \frac{(1-e)^{3/2}}{(1+e)^{1/2}}<c<1-e , 
\nonumber\\
&&c^{4/3}|1-e|^{2/3}-c^2 \leq R \leq (1-e)^2-c^2 , 
\\
&&c^{4/3}(1+e)^{2/3}-c^2 \leq R \leq (1+e)^2-c^2 , 
\\
\mbox{for} && 1-e<c<1+e , 
\nonumber\\
&&c^{4/3}(1+e)^{2/3}-c^2 \leq R \leq (1+e)^2-c^2 , 
\\
\mbox{for} && 1+e<c , 
\nonumber\\
&&\mbox{no allowed region for} \, R. 
\label{region2}
\end{eqnarray}

These cases show that there are two critical curves 
for a small core radius $c<1-e$, 
one for the quite large case $1-e<c<1+e$, 
and they disappear for an extremely large one, $1+e<c$. 
This anomalous behavior where the critical curves 
disappear is physically understood as follows. 
As the core radius becomes much larger, the lens system 
approaches an approximately constant mass sheet.  
Around $c\sim 1+e$, the size of the core is comparable 
to the Einstein ring radius. 
As a result, such a lens system cannot produce multiple 
images, which means that there are no critical curves. 
Furthermore, the latter two cases of large cores are not realistic. 
Indeed, $c\sim 1$ implies that the core radius is comparable 
to the Einstein ring radius, which becomes about 10kpc
for a galactic lens with $10^{12}$ solar mass at a high redshift. 
The core radius of a galaxy is usually much smaller 
than 10kpc. 

It is convenient to consider Eq. ($\ref{critical2}$) 
as an $r$-parameter representation of the critical curves, 
because we must solve Eq. ($\ref{critical2}$) with respect to 
$r$ to obtain a $\phi$-parameter representation. 
This $r$-representation becomes 
\begin{equation}
(x, y)=(\pm r\sqrt{\frac{1+h}{2}}, \pm r\sqrt{\frac{1-h}{2}}) , 
\label{critical-rep}
\end{equation}
where we defined 
\begin{eqnarray}
h=(eR)^{-1} 
&\Bigl(&(R+c^2)^{3/2}-(R+2c^2)
\nonumber\\
&&
+c^2(1-e^2)(R+c^2)^{-1/2}\Bigr) . 
\label{h}
\end{eqnarray}
The allowed regions for $r=\sqrt{R}$ are given by 
Eqs. ($\ref{region1}$)-($\ref{region2}$) depending on $c$ and $e$. 
Substituting this representation into Eqs. ($\ref{lenseq-cie1}$) 
and ($\ref{lenseq-cie2}$), we obtain a representation of 
the caustics as 
\begin{eqnarray}
a(r)&=&\pm r\Bigl(1-\frac{1-e}{\sqrt{r^2+c^2}}\Bigr)
\sqrt{\frac{1+h}{2}} , 
\label{caustics-rep1}
\\
b(r)&=&\pm r\Bigl(1-\frac{1+e}{\sqrt{r^2+c^2}}\Bigr)
\sqrt{\frac{1-h}{2}} . 
\label{caustics-rep2}
\end{eqnarray}

As an illustration, the caustics and critical curves are shown 
for $e=0.1$ and $c=0.2$ in Figs. 1 and 2, respectively. 
We can see two curves in both figures as discussed above. 
The inner caustics in Fig. 1 look like an asteroid. 
However, it is not exact but approximate, since 
Eqs. ($\ref{caustics-rep1}$) and ($\ref{caustics-rep2}$) 
are not the equation for asteroids (Asada et al. 2003). 

\begin{figure}
\centering
\includegraphics[width=8cm]{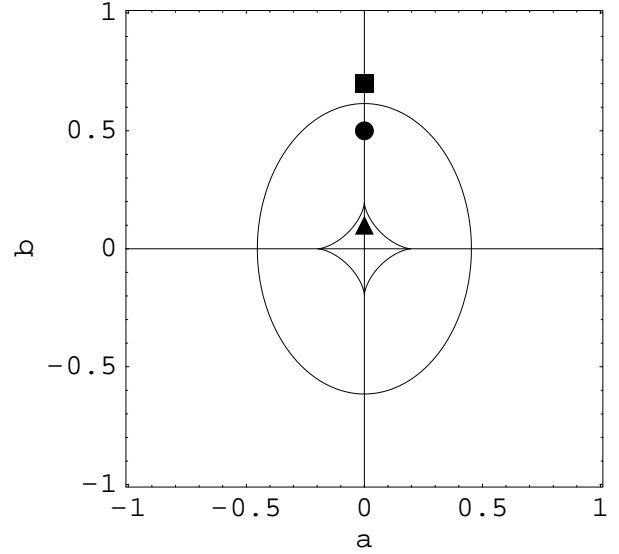}
\caption{
Caustics for a cored isothermal ellipsoidal lens 
for $e=0.1$ and $c=0.2$. 
Sources are located at $(0, 0.1)$, $(0, 0.5)$ and $(0, 0.7)$, 
denoted by the triangle, filled disk and square, respectively. 
}
\label{Fig1}
\end{figure}

\begin{figure}
\centering
\includegraphics[width=8cm]{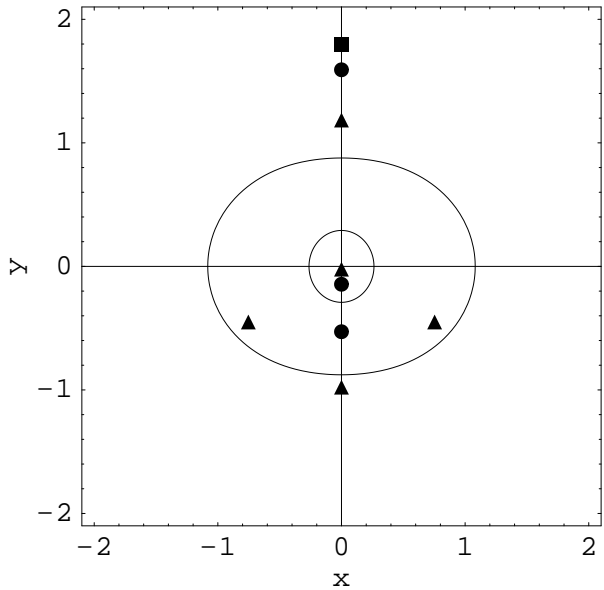}
\caption{
Critical curves for a cored isothermal ellipsoidal lens 
for $e=0.1$ and $c=0.2$. 
The images correspond to the sources in Fig. 1. 
}
\label{Fig2}
\end{figure}

\section{Conclusion}
We have studied the effect of the core radius of 
an isothermal ellipsoid on the caustics and critical curves. 
For this purpose, we have obtained a simple parametric 
representation of the critical curves. 
The caustics have been expressed also with the same parameter. 
It turns out that a nonvanishing core radius changes 
the shape and number of the caustics and critical curves.  
These curves are 
two for a small core radius $c<1-e$, 
one for a quite large case $1-e<c<1+e$, 
and disappear for an extremely large one $1+e<c$, 
where $c$ is the angular core radius and 
$e$ is the ellipticity of the isothermal ellipsoid. 
However, the latter two cases are not realistic 
in astronomical situations as discussed above. 

We must perform numerical computations for various values 
of the parameters in order to obtain the caustics and critical curves. 
The present expression is analytic. 
Hence, it enables us to save time and reach more accurate
results, for instance in rapid and accurate estimations 
of a lensing event rate, fittings to observational data, etc.

\begin{acknowledgements}
The author would like to thank M. Bartelmann, Y. Mellier 
and M. Kasai for useful conversations. 
He would like to thank L. Blanchet for hospitality at the Institute 
for Astrophysics at Paris. 
This work was supported by a fellowship for visiting scholars from 
the Ministry of Education of Japan. 
\end{acknowledgements}

\end{document}